\documentclass[conference]{IEEEtran}
\IEEEoverridecommandlockouts
\usepackage{cite}
\usepackage{amsmath,amssymb,amsfonts}
\usepackage{graphicx}
\usepackage{textcomp}
\usepackage{xcolor}
\def\BibTeX{{\rm B\kern-.05em{\sc i\kern-.025em b}\kern-.08em
    T\kern-.1667em\lower.7ex\hbox{E}\kern-.125emX}}

\usepackage[frozencache ,cachedir=.]{minted}

\usepackage{svg}
\usepackage{minted}
\usepackage{tcolorbox}
\usepackage{enumitem}
\usepackage{lipsum}
\usepackage{xcolor, soul}
\usepackage{float}
\usepackage{dblfloatfix}

\usepackage{algorithm} 
\usepackage[noend]{algpseudocode}

\usepackage{multicol}

\makeatletter
\newcommand{\newlineauthors}{%
  \end{@IEEEauthorhalign}\hfill\mbox{}\par
  \mbox{}\hfill\begin{@IEEEauthorhalign}
}
\makeatother

\begin{document}

\title{Detecting Unknown Attacks in IoT Environments: An Open Set Classifier for Enhanced Network Intrusion Detection
}
\makeatletter
\patchcmd{\@maketitle}
  {\addvspace{0.5\baselineskip}\egroup}
  {\addvspace{-1.5\baselineskip}\egroup}
  {}
  {}
\makeatother




\author{\IEEEauthorblockN{Yasir Ali Farrukh}
\IEEEauthorblockA{\textit{Clean and Resilient Energy Systems (CARES) Lab}\\
\textit{Texas A\&M University} \\
College Station, TX, USA \\
yasir.ali@tamu.edu}\\
\IEEEauthorblockN{Irfan Khan}
\IEEEauthorblockA{\textit{Clean and Resilient Energy Systems (CARES) Lab}\\
\textit{Texas A\&M University} \\
Galveston, TX, USA \\
irfankhan@tamu.edu}\\
\and
\IEEEauthorblockN{Syed Wali}
\IEEEauthorblockA{\textit{Clean and Resilient Energy Systems (CARES) Lab}\\
\textit{Texas A\&M University} \\
College Station, TX, USA \\
syedwali@tamu.edu}\\

\IEEEauthorblockN{Nathaniel D. Bastian}
\IEEEauthorblockA{\textit{Army Cyber Institute}\\
\textit{United States Military Academy} \\
West Point, NY, USA \\
nathaniel.bastian@westpoint.edu}
}

\maketitle

\begin{abstract}
The widespread integration of Internet of Things (IoT) devices across all facets of life has ushered in an era of interconnectedness, creating new avenues for cybersecurity challenges and underscoring the need for robust intrusion detection systems. However, traditional security systems are designed with a closed-world perspective and often face challenges in dealing with the ever-evolving threat landscape, where new and unfamiliar attacks are constantly emerging. In this paper, we introduce a framework aimed at mitigating the open set recognition (OSR) problem in the realm of Network Intrusion Detection Systems (NIDS) tailored for IoT environments. Our framework capitalizes on image-based representations of packet-level data, extracting spatial and temporal patterns from network traffic. Additionally, we integrate stacking and sub-clustering techniques, enabling the identification of unknown attacks by effectively modeling the complex and diverse nature of benign behavior. The empirical results prominently underscore the framework's efficacy, boasting an impressive 88\% detection rate for previously unseen attacks when compared against existing approaches and recent advancements. Future work will perform extensive experimentation across various openness levels and attack scenarios, further strengthening the adaptability and performance of our proposed solution in safeguarding IoT environments.
\end{abstract}



\section{Introduction}

The rapid proliferation of Internet of Things (IoT) devices has directed a new era of interconnectedness, revolutionizing various sectors like healthcare, transportation, agriculture, other industries \cite{li2018learning}, and the military. These IoT ecosystems consist of interconnected sensors, actuators, and network-enabled devices, facilitating data exchange through the internet \cite{bhattacharya2020novel}. However, the exponential growth of IoT systems, projected to reach 75.3 billion devices by 2025 \cite{akhgar2015application}, has also introduced new avenues for cyberattacks, posing significant challenges to the security and privacy of interconnected devices and their data. As adversaries become more sophisticated, traditional security measures like Network Intrusion Detection Systems (NIDS) relying on closed-world settings \cite{close_world} face unprecedented challenges in safeguarding IoT environments. Such NIDS are tested only against known attack classes, rendering them ineffective against previously unseen attacks. In contrast, effective security solutions must address open-world network intrusion detection settings, where classifiers must detect unknown attack classes. These types of classifiers are known as open-set classifiers, while those relying on closed-world settings are termed close-set classifiers \cite{open_set}.

As the boundaries between benign and malicious behaviors blur \cite{zhangdetecting}, there is an urgent need for a more robust and proactive security approach that can accurately identify unknown/novel attacks in real-time, effectively mitigating their impact on IoT systems. In response to this challenge, our paper introduces an innovative framework for an open-set classifier tailored to IoT devices in adversarial environments. The framework utilizes the stacking concept \cite{stacking} and diverse prototypical features of benign traffic to spot deviations from normal behavior. It classifies incoming network traffic as either benign or unknown attacks. By adopting an open-set problem formulation, our approach confidently distinguishes between benign traffic and entirely new threats, even without prior training data. 

Our contributions encompass not only the proposal of an open-set classifier tailored for IoT environments but also a different approach to utilizing the network traffic of IoT devices as serialized RGB images. Unlike traditional closed-set classifiers that rely on flow-based data, our open-set classifier operates at the packet level of IoT network traffic. This granular approach allows us to easily distinguish novel attacks, as flow-based data lacks the actual message content of each flow. In addition to our contributions, we have conducted a thorough evaluation of our approach against diverse attack scenarios, demonstrating its efficacy in detecting and accurately classifying unseen threats. The experimental validation showcases the superiority of our approach over traditional closed-set NIDS and state-of-the-art open-set classifiers, highlighting its potential to enhance IoT security significantly.

\section{Related Works For Open-set Classification}

Efforts in anomaly detection within NIDS have been extensive, aiming to differentiate normal network traffic from malicious patterns. However, a substantial portion of this work predominantly addresses the closed-world problem, where models are designed to recognize only the classes encountered during training. This presents a challenge when models need to identify classes not seen during training, constituting the open-set recognition (OSR) problem \cite{baye2023performance}.

Pioneers in the pursuit of OSR, Scheirer et al., formally defined the problem \cite{scheirer2012toward}. They introduced a pioneering 1-vs-set machine-aided solution, followed by the innovative Compact Abating Probability (CAP). A notable instantiation within CAP is the W-SVM, utilizing Statistical Extreme Value Theory (EVT) to calibrate SVM decision scores. The efficacy of W-SVM is demonstrated by Cruz for fine-grained open-set intrusion detection \cite{WSVM}. Further, Chen et al. proposed an Auto-encoder Ensemble \cite{AE} approach exploiting the variable connectivity architecture of auto-encoders for improved performance, while Bradley et al. leveraged survival analysis \cite{10207602}.

In recent advancements, Ruff et al. introduced DeepSAD \cite{DeepSAD}, a semi-supervised approach grounded in the idea that the entropy of the latent distribution for normal data should exhibit lower values than that of anomalous samples. Pang et al. presented PreNet \cite{PreNet}, a novel deep weakly-supervised approach focused on learning pairwise relation features and anomaly scores through predicting relationships between randomly sampled instances. Li et al. proposed the ECOD \cite{ECOD} algorithm, inspired by outliers often being ``rare events" in the tails of a distribution. Despite these developments, the NIDS domain has seen few works directly addressing the OSR challenge. Baye et al. recently conducted an empirical study, exploring notable OSR algorithms using NIDS data to uncover correlations between deep learning-based OSR algorithms' performance and hyperparameter values they use \cite{baye2023performance}.

In sum, the efficacy of NIDS in an open-world context is limited as most machine learning-based NIDS operate within a closed-world setting \cite{close_world}. This underscores the pressing need for further progress and innovation in this field.

\section{Methodology}

This section offers a comprehensive overview of our methodology, commencing with the dataset employed in our experiments. Subsequently, we delve into the preprocessing steps undertaken to ready the data for training and testing. Additionally, we elaborate on the clustering procedure applied to the serialized RGB network traffic images to determine the optimal number of clusters, denoted as $N$. Lastly, we provide an in-depth description of our proposed framework.

\subsection{Dataset and Preprocessing}

The dataset used to evaluate our proposed framework is CIC-IDS2017 \cite{Sharafaldin2018TowardCharacterization}, created by the University of New Brunswick in 2017. It consists of simulated network traffic in both packet-based and bidirectional flow-based formats, encompassing the most up-to-date attacks and benign traffic. The dataset is available in two formats: the original packet capture (PCAP) files (packet-based data) and CSV files (flow-based data) obtained by extracting 80 features from the PCAPs using CICFlowMeter. 

For our work, we specifically used the packet-based data from CIC-IDS2017, as flow-based data cannot detect attacks that rely on packet's payload \cite{ullah2020two}. Additionally, packet-based data of CIC-IDS2017 is not labeled; therefore, we first labeled the data utilizing our developed tool (Payload-Byte) \cite{farrukh2022payload} that extracts and labels packet capture files of network traffic using metadata from NIDS datasets. The tool leverages five-tuple features, including Source IP, Destination IP, Source Port, Destination Port, and Protocol, to match packets with labeled flow-based data instances. The resulting labeled features are payload content. Since payload size varies for each packet, Payload-Byte uses a maximum payload length of 1500 bytes. The extracted payload forms one large feature, which is then divided into 1500 features based on bytes. Each byte's hexadecimal representation is transformed into an integer ranging from 0 to 255, resulting in one feature. For packets with fewer than 1500 payload bytes, zero padding is employed to maintain a standardized feature vector structure. After labeling the data, we removed duplicated instances and instances with no payload data. Furthermore, we performed under-sampling to reduce the dataset size by decreasing the number of benign instances. For a comprehensive understanding of the preprocessing steps and the functioning of Payload-Byte, we refer readers to our previous work \cite{farrukh2022payload}.

\begin{figure}[htbp]
  \centering
  \includegraphics[scale=0.65]{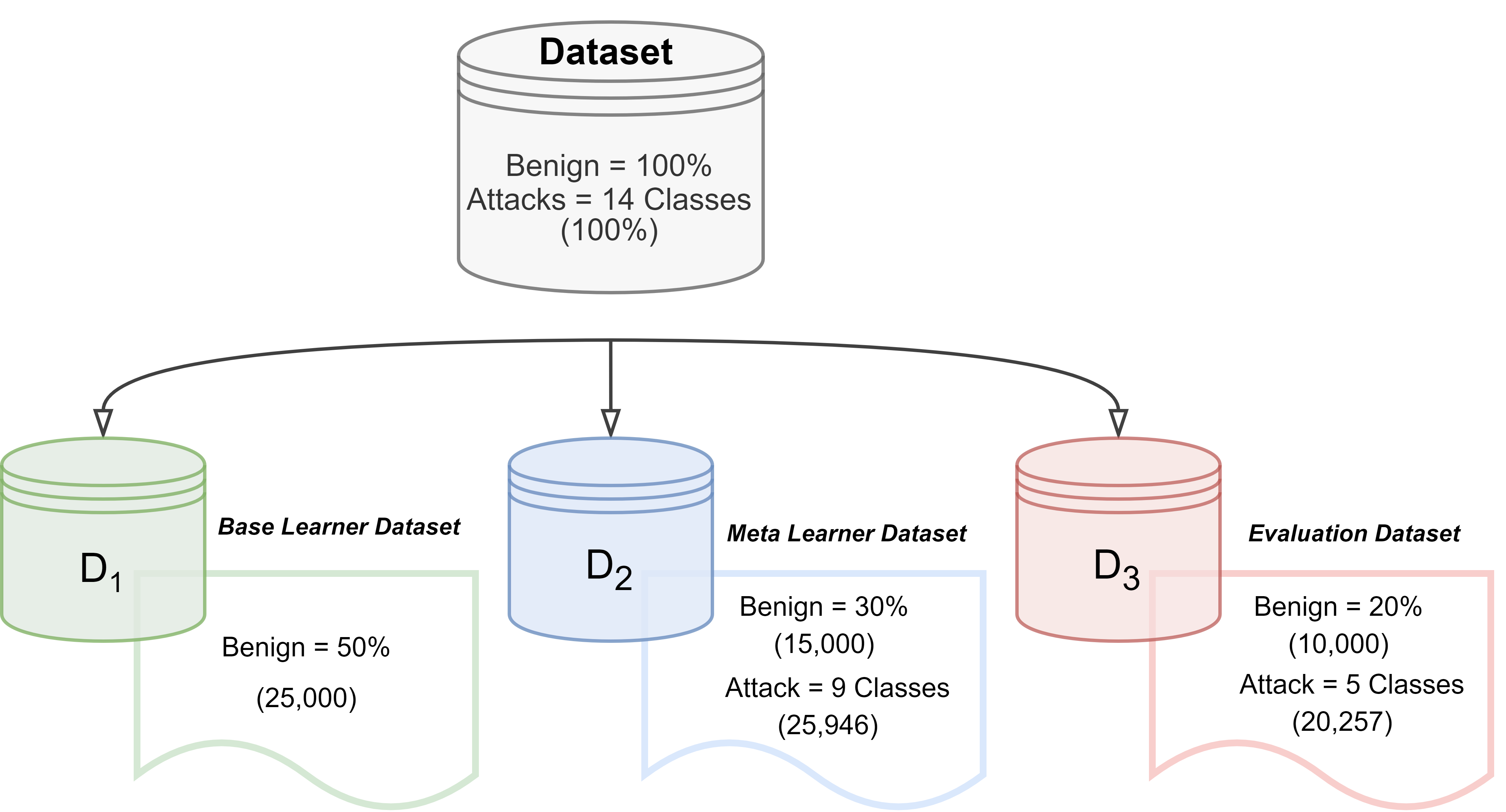}
  \caption{Illustration of dataset division into three sub-datasets. The sub-dataset ($D_3$) includes five randomly separated attack classes, while the rest of the attack classes are included in $D_2$. On the other hand, $D_1$ exclusively consists of benign data.}
  \label{fig:data_div}
\end{figure}

\begin{figure*}[!htbp]
  \centering
  \includegraphics[scale=0.62]{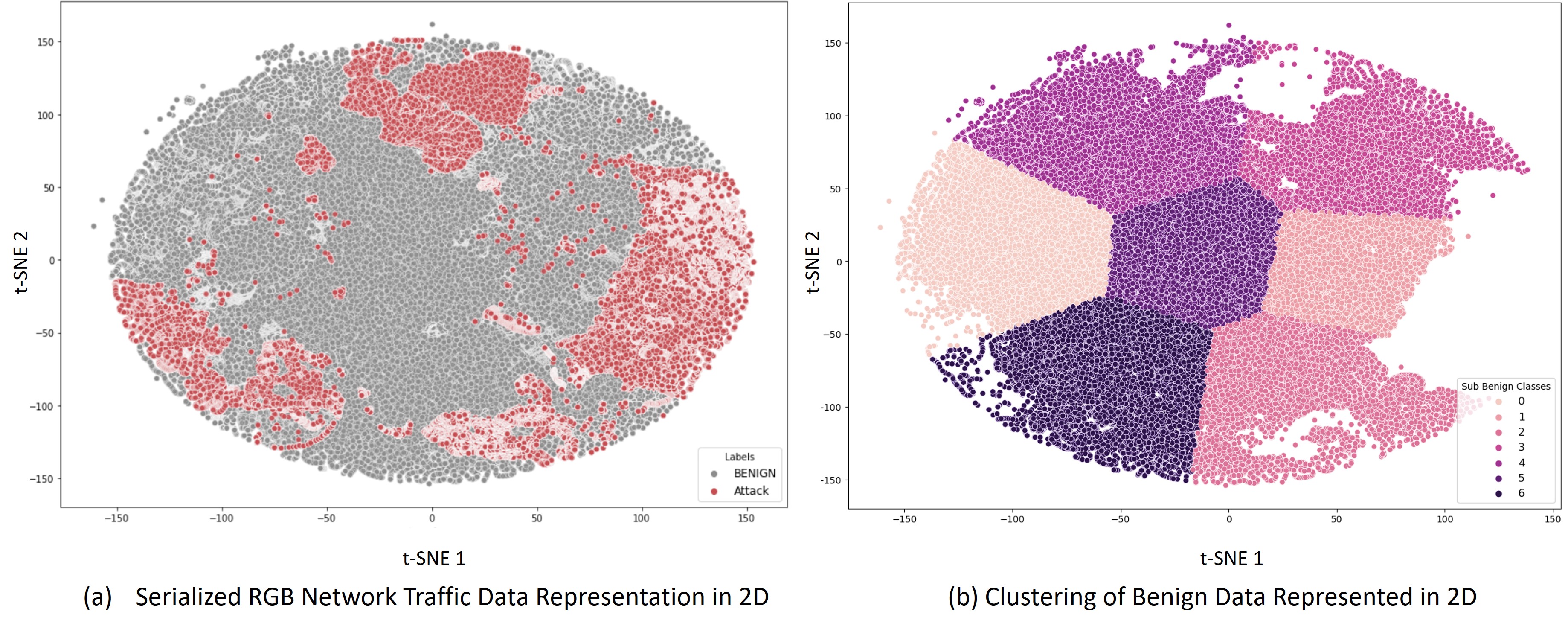}
  \caption{Representation of Serialized RGB images of network traffic into two dimensions using t-SNE method. (a) illustrate the distribution of benign data and attacks, highlighting the diverse nature. (b) provides an insightful depiction of the effective clustering of the benign data into seven distinct clusters.}
  \label{fig:cluster}
\end{figure*}

After labeling and preprocessing the data, we converted it into serialized RGB images, following the methodology used in our previous work \cite{farrukh4370422senet}. After the transformation, the data was divided into three sub-datasets:

\begin{itemize}
\item \textbf{Base Learner Dataset ($D_1$):} This dataset exclusively contains benign data and serves as the training set for the base learner models.
\item \textbf{Meta Learner Dataset ($D_2$)}: Comprising nine known attack classes and benign data samples. This dataset is utilized for generating meta features and training the meta classifier.
\item \textbf{Evaluation Dataset ($D_3$)}: This dataset forms the testing dataset, consisting of five unknown attacks and benign samples. 
\end{itemize} 

Our primary aim of evaluating the OSR problem for NIDS involves detecting unknown classes without prior knowledge. To achieve this, we separated five random attack classes \textit{(DoS Hulk, DoS slowloris, DoS Slowhttptest, Web Attack–Sql Injection and Bot)} from the dataset to generate unknown attack scenarios. Furthermore, we partitioned the benign data samples in a ratio of 50:30:20 for the base learner, meta learner, and evaluation datasets, respectively. The remaining nine attack classes were treated as known attacks and included in meta learner dataset along with 30\% of benign data. The complete distribution of the dataset is illustrated in Fig. \ref{fig:data_div}.

\subsection{Clustering of Benign Network Traffic}

Incorporating the clustering of benign traffic within our framework offers a crucial enhancement to the performance of the proposed framework. As benign data is usually more spread out in the feature space than attack data, this makes it difficult to distinguish between normal instances and unknown attacks \cite{cluster}. By dividing the benign data into sub-clusters, we aim to capture inherent variations and nuances in benign behavior patterns, thereby facilitating a more nuanced and accurate classification. This stratification enables our framework to differentiate between benign  and unknown attacks with higher precision, contributing to a reduced rate of false positives. 

\begin{figure}[htbp]
  \centering
  \includegraphics[scale=0.52]{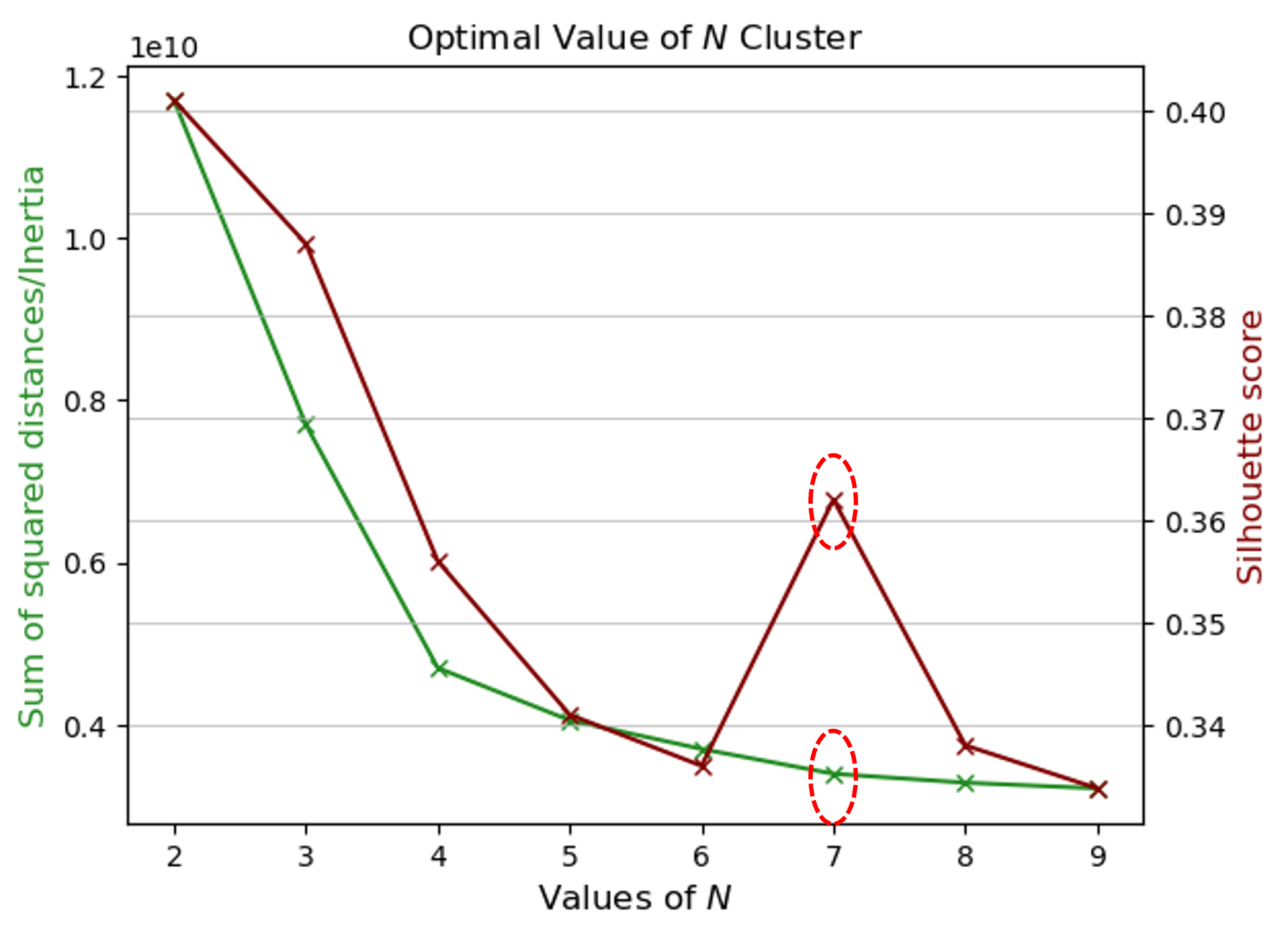}
  \caption{Graph for sum of squared distances and silhouette score for different number of clusters. Green line represents the values obtained through elbow method, and the maroon line represents the silhouette score. The optimal number of clusters for benign is found to be seven, shown by red circles.}
  \label{fig:elbow}
\end{figure}

\begin{figure*}[!htbp]
  \centering
  \includegraphics[scale=0.62]{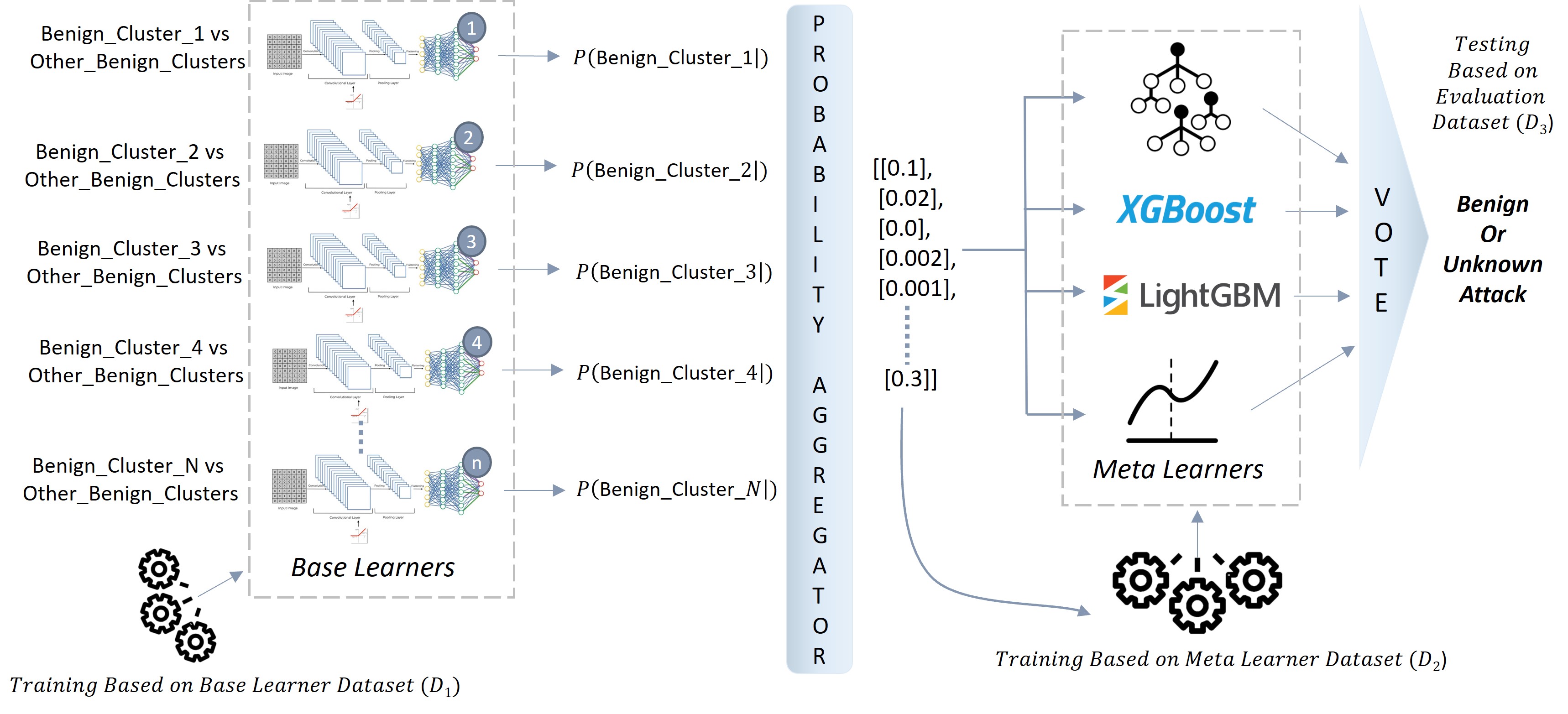}
  \caption{Pictorial representation of the proposed framework for detecting unknown attacks in IoT environments. The framework consists of two levels: Base Learner Models and Meta Learner Models. Each level is trained using a different subset of the dataset.}
  \label{fig:framework}
\end{figure*}

Initially, we converted the transformed (Serialized RGB Images) dataset into a two-dimensional space using the t-distributed Stochastic Neighbor Embedding (t-SNE) method \cite{van2008visualizing}. The two-dimensional space representation of our transformed data can be seen in Fig. \ref{fig:cluster}. Notably, the benign data displays a dispersed distribution across the space, while instances of attacks overlap with it, posing a challenge in distinguishing between benign and attack instances. Subsequently, this two-dimensional representation of the data formed the basis for both visualization and the subsequent clustering process of benign traffic using K-means clustering.

To determine the optimal number of clusters ($N$), we employed two widely used methods in the literature: the Elbow method and the Silhouette method \cite{bohara2020survey}. These techniques aid in identifying the most suitable value of $N$, which is fundamental for effective clustering. The Elbow method focuses on the point where the reduction in within-cluster variance starts to slow down, indicating an appropriate number of clusters. The Silhouette method, on the other hand, assesses the quality of the clustering based on cohesion and separation of clusters. 

Fig. \ref{fig:elbow} showcases the outcomes yielded by the Elbow and Silhouette methods. These graphs show that the optimal number of clusters, denoted as $N$, is seven. This determination resonates in our framework, resulting in the adoption of seven base-learner models that are discussed in the subsequent heading. The two-dimensional projection of the benign data as well as how benign data is clustered into seven sub-benign clusters, is visually represented in Fig. \ref{fig:cluster}. After successful clustering, we leverage the resultant cluster labels to annotate the sub-clusters of benign data in the serialized RGB data format which is utilized for training the base-learner models.


\begin{algorithm*}
\caption{Proposed Framework for Detecting Unknown Attacks in IoT Environments}
\label{algorithm:framework}
\begin{algorithmic}[1]
\Statex \textbf{Input:} Benign data, Known and Unknown attack data
\Statex \textbf{Output:} Prediction $\rightarrow $ Benign or Unknown Attack
\State \textbf{Step 1: Dataset Division}
\State \textit{$D_1:$ Base Learner Dataset} $\leftarrow $ Benign data
\State \textit{$D_2:$ Meta Learner Dataset} $\leftarrow $ Benign + Known attacks data
\State \textit{$D_3:$ Evaluation Dataset} $\leftarrow $ Benign + Unknown attacks data
\State \textbf{Step 2: Sub-Clustering of Benign Data}
\State Determine optimal number $N$ of benign clusters using Elbow and Silhouette methods
\State Apply unsupervised K-means clustering to divide the benign data of \textit{$D_1$} into $N$ sub-classes, 
\For{$i = 1$ to $N$}
    \State \textbf{Step 3: Base Learner Model Training}
    \State Train base learner models on the data of $i$-th benign cluster against the rest of the benign data
\EndFor
\State \textbf{Step 4: Meta Learner Model Training}
\State Pass \textit{$D_2$} through the trained Base Learner Models to obtain probabilities $P_i$
\State Aggregate the probabilities $P_i$ from each sample across all Base Learner Models to generate meta features.
\State Train meta-classifiers (Random Forest, Logistic Regression, XGBoost, and LightLGBM) on the combined meta features
\State \textbf{Step 5: Testing}
\State Feed \textit{$D_3$} through trained Base and Meta Learner Models to obtain predictions
\State Implement majority voting among the Meta Learner Models' outputs to finalize the prediction
\State $V = \frac{1}{|M|} \sum_{i=1}^{|M|} O_i$ \Comment{$M$ = Set of meta-
classifiers; $O_i$ = Output of $i$-th meta-classifier}
\If {\(V \geq 0.5\)}
\State \textbf{Output} $\rightarrow $ Unknown Attack
\Else
\State \textbf{Output} $\rightarrow $ Benign
\EndIf

\end{algorithmic}
\end{algorithm*}

\subsection{Framework}

Our proposed framework builds upon our previous work \cite{farrukh4370422senet}, which initially focused on the closed-set classifier approach. While the earlier work provided preliminary insights, this paper extends the methodology to encompass open-set scenarios. A visual depiction of our proposed framework is presented in Fig. \ref{fig:framework}. In the context of detecting unknown attacks within IoT environments, our framework draws inspiration from the concept of Stacking \cite{stack_1,stack_2}, which is a Meta Learning based modeling technique consisting of two types of learners: Base Learners and Meta Learners.

For the base learners, we build upon the architecture used in our previous work \cite{farrukh4370422senet}, which involves a deep concatenated Convolutional Neural Network (CNN). Notably, the base learners are solely trained on benign data utilizing $D_1$ (sub-dataset). Given the diverse nature of benign behavior patterns, distinguishing benign data as a whole from novel attacks becomes challenging. To address this, we adopt an unsupervised clustering method, K-means, to divide the benign data into $N$ sub-classes. We then train $N$ base learner models, each based on binary classification, to discern whether a data sample belongs to its particular benign cluster or not.

In other words, we train each model to distinguish samples from its specific cluster versus the rest of the benign clusters. Consequently, after training the base learners, we obtain $N$ probabilities indicating the likelihood that a given sample belongs to each respective cluster. This approach allows us to gain insights into the association of a sample with each sub-class of benign behavior, aiding in the accurate detection and classification of novel attacks.

Next, we utilize $D_2$ subset of the dataset and feed it through the base learners, producing meta features based on the $N$ probabilities from each model. These meta features are then used to train the meta-classifiers, which include Random Forest, Logistic Regression, XGBoost, and LightLGBM.

Once the meta-classifiers are trained, the training process of our framework is completed, and we can evaluate its performance using $D_3$ (sub-dataset) . Since there are four meta classifiers, we obtain four outputs indicating whether a sample is benign or an unknown attack. To mitigate potential conflicts in the outputs, we incorporate a voting ensemble mechanism. 

Let $M$ be the set of meta-classifiers, where $|M|$ represents the total number of meta-classifiers. Each meta-classifier $m_i \in M$ produces an output $O_i$ for a given input sample. The outputs can be binary, where $O_i=1$ indicates a predicted attack, and $O_i=0$ indicates a predicted benign sample. The voting mechanism is implemented as follows:

\begin{equation}
V = \frac{1}{|M|} \sum_{i=1}^{|M|} O_i
\end{equation}

\noindent where \(V\) is the final voting result. If \(V \geq 0.5\), the sample is classified as an attack; otherwise, it is considered benign.


The overall training and testing process of our framework is detailed in Algorithm \ref{algorithm:framework}. This algorithm outlines the sequential steps, from training the base learners to combining meta features and utilizing meta classifiers for making predictions.

\section{Results and Discussion}

To evaluate our proposed framework, we compare the performance with several novelty and out-of-distribution detection approaches, as well as state-of-the-art methods in detecting unknown attacks to comprehensively assess its effectiveness. To ensure an equitable comparison, we ensure that each approach is evaluated under analogous experimental setting, utilizing the same packet-level dataset and similar data division. 

\begin{figure}[htbp]
  \centering
  \includegraphics[width=\linewidth]{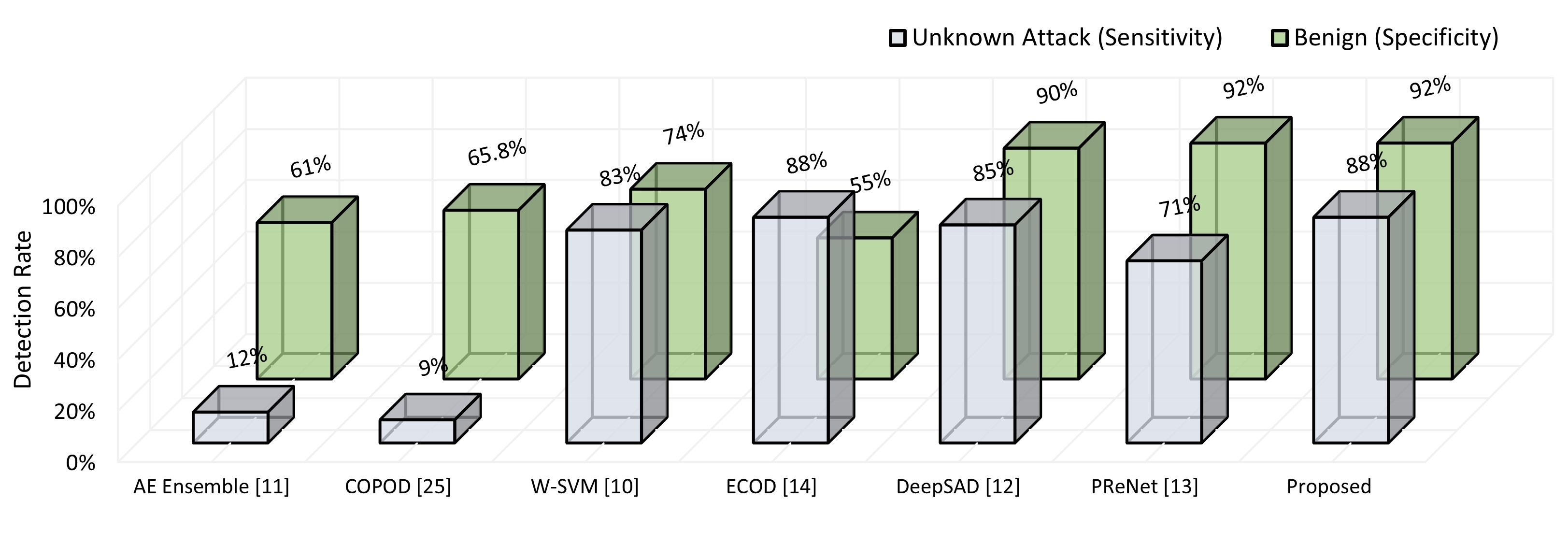}
  \caption{Comparison of the detection rates of unknown attacks and benign samples with other approaches. The proposed framework outperforms other available approaches in terms of detecting unknown attacks.}
  \label{fig:result}
\end{figure}

Our experimental evaluation aimed to assess the effectiveness of our proposed framework by considering the detection rate of unknown attacks (sensitivity/recall) and the detection rate of benign samples (specificity) as our evaluation metrics. The summarized results can be observed in Fig. \ref{fig:result}, which provides an overview of the performance metrics for each approach.

From the figure, it is evident that our approach closely matches the sensitivity performance of the ECOD approach. However, noteworthy differences arise in terms of specificity, where our approach outperforms ECOD by 37\%. This significant discrepancy indicates our approach's superiority in this aspect. Similarly, our framework achieves comparable specificity with the PReNet approach but simultaneously leads by 17\% in sensitivity. Overall, our framework exhibits comprehensive performance when it comes to detecting both unknown attacks and benign behaviors. A crucial aspect of an open-set classifier is balancing specificity and sensitivity, a challenge our proposed framework adeptly manages.

The superiority of our proposed framework over other approaches can be attributed to three main factors. Firstly, the utilization of packet-based data and its image-based representation enables the extraction of both spatial and temporal information from network traffic. This empowers our framework to identify subtle patterns and anomalies within the data, significantly enhancing its ability to distinguish between unknown attacks and normal traffic. Secondly, the subdivision of benign data contributes to a clearer depiction of the inherent data distributions. This division aids in capturing the intricate and varied patterns intrinsic to benign behavior. As a result, our framework excels in detecting previously unseen attacks by comprehensively understanding the complex behaviors present within the benign class.

\section{Conclusion and Future Work}

In this paper, we present a novel framework designed specifically for open-set classification within the domain of NIDS in adversarial IoT environments. The key innovation of our framework resides in its utilization of packet-level data, which is transformed into serialized RGB images. This distinctive approach enables us to harness both the spatial and temporal information inherent in the network traffic data, providing a richer and more comprehensive understanding of the underlying patterns.

By combining the principles of stacking and sub-clustering within our framework, we effectively address the intricate challenge of identifying unknown attacks amidst the ever-evolving cybersecurity landscape. Our experimental findings underline the remarkable efficacy of our framework, boasting an impressive 88\% detection rate for previously unseen attacks that were not encountered during the training phase. It is important to note that this paper lays the foundation for our proposed framework, with comprehensive experimentation and evaluation across varying degrees of openness and attack scenarios forming a significant part of our future work. Through these continued efforts, we aim to further validate and fine-tune the capabilities of our framework to provide enhanced capability for NIDS in adversarial IoT environments.


\section*{Acknowledgment}

This work was supported in part by the U.S. Military Academy (USMA) under Cooperative Agreement No. W911NF-22-2-0081, the U.S. Army Combat Capabilities Development Command (DEVCOM) Army Research Laboratory under Support Agreement No. USMA 21050, and the U.S. Army DEVCOM C5ISR Center under Support Agreement No. USMA21056. The views and conclusions expressed in this paper are those of the authors and do not reflect the official policy or position of the U.S. Military Academy, U.S. Army, U.S. Department of Defense, or U.S. Government.

Research reported in this paper was also supported by an Early-Career Research Fellowship from the Gulf Research Program of the National Academies of
Sciences, Engineering, and Medicine. The content is solely the responsibility of the authors and does not
necessarily represent the official views of the Gulf Research Program of the National Academies
of Sciences, Engineering, and Medicine.


%


\nocite{COPOD}

\bibliographystyle{IEEEtran}
\bibliography{references.bib}

\end{document}